\documentclass{article}

\usepackage{microtype}
\usepackage{graphicx}
\usepackage{epsfig}
\usepackage{booktabs} % for professional tables
\usepackage{amsmath}
\usepackage{amssymb}
\usepackage{bm}
\usepackage{color}
\usepackage[table,xcdraw]{xcolor}
\usepackage{algorithm}
\usepackage{algorithmic}
\usepackage[labelformat=simple]{subcaption}
\usepackage{multirow}
\usepackage{graphbox} %loads graphicx package
\usepackage{array}

\newcolumntype{C}[1]{>{\centering\arraybackslash}p{#1}}

\newcommand{\eg}{\textit{e.g. }}
\newcommand{\ie}{\textit{i.e. }}

\newenvironment{packed_enum}{
\begin{enumerate}
  \setlength{\itemsep}{1pt}
  \setlength{\parskip}{0pt}
  \setlength{\parsep}{0pt}
}{\end{enumerate}}

\usepackage{hyperref}

\usepackage[accepted]{icml2020}

\icmltitlerunning{Tuning-free Plug-and-Play Proximal Algorithm for Inverse Imaging Problems}

\begin{document}

\twocolumn[
\icmltitle{Tuning-free Plug-and-Play Proximal Algorithm for Inverse Imaging Problems}




\begin{icmlauthorlist}
\icmlauthor{Kaixuan Wei}{BIT}
\icmlauthor{Angelica Aviles-Rivero}{Cambridge1}
\icmlauthor{Jingwei Liang}{Cambridge2}
\icmlauthor{Ying Fu}{BIT}
\icmlauthor{Carola-Bibiane Sch\"{o}nlieb}{Cambridge2}
\icmlauthor{Hua Huang}{BIT}
\end{icmlauthorlist}

\icmlaffiliation{BIT}{School of Computer Science and Technology, Beijing Institute of Technology, Beijing, China}
\icmlaffiliation{Cambridge1}{DPMMS, University of Cambridge, Cambridge, United Kingdom}
\icmlaffiliation{Cambridge2}{DAMTP, University of Cambridge, Cambridge, United Kingdom}
\icmlcorrespondingauthor{Ying Fu}{fuying@bit.edu.cn}

\icmlkeywords{Machine Learning, ICML}

\vskip 0.3in
]



\printAffiliationsAndNotice{}  % leave blank if no need to mention equal contribution

\begin{abstract}
Plug-and-play (PnP) is a non-convex framework 
that combines ADMM or other proximal algorithms  with advanced denoiser priors.  
Recently, PnP has achieved great empirical  success, especially with the integration of deep learning-based denoisers. 
However, a key problem of PnP based approaches is that they require manual parameter tweaking. It is necessary to obtain high-quality results across the high discrepancy in terms of imaging conditions and varying scene content.
In this work, we present a tuning-free PnP proximal algorithm, which can automatically  determine the internal parameters including  the penalty parameter, the denoising strength  and the terminal time.
A key part of our approach is to develop a policy network for automatic search of parameters, which can be effectively learned via mixed model-free and model-based deep reinforcement learning. 
We demonstrate, through numerical and visual experiments, that the learned policy can customize different parameters for different states, and often more efficient and effective than existing handcrafted criteria. Moreover, we discuss the practical considerations of the plugged denoisers, which together with our learned policy yield  state-of-the-art results. This is prevalent on both linear and nonlinear exemplary inverse imaging problems, and in particular, we show promising results on Compressed Sensing MRI and phase retrieval. 
\end{abstract}

\section{Introduction}
The problem of recovering an underlying unknown image $x \in \mathbb{R}^N$ from noisy and/or incomplete measured data $y \in \mathbb{R}^M$ is fundamental in computational imaging, in applications including magnetic resonance imaging (MRI) \cite{fessler2010model}, computed tomography (CT) \cite{elbakri2002segmentation}, microscopy \cite{aguet2008model,zheng2013wide}, and inverse scattering \cite{katz2014non,metzler2017coherent}, to name a few. This image recovery task is often formulated as an optimization problem that minimizes a cost function, i.e., 
%as an inverse problem which is often underdetermined and highly ill-posed.  
%Generally, ill-posed problems are solved by minimizing a cost functional 
 \begin{align}
 \mathop{\mathrm{minimize}}_{x \in \mathbb{R}^N} \quad \mathcal{D} \left( x \right)+ \lambda \mathcal{R} \left( x \right),
\label{eq:optimization}
\end{align}
where $\mathcal{D}$ is a data-fidelity term that ensures consistency between the reconstructed image and measured data. $\mathcal{R}$ is a regularizer that imposes certain prior knowledge, \eg smoothness \cite{osher2005iterative,ma2008efficient}, sparsity \cite{yang2010fast,liao2008sparse,ravishankar2010mr}, low rank \cite{semerci2014tensor,gu2017weighted} and nonlocal self-similarity \cite{mairal2009non,qu2014magnetic}, regarding the unknown image.  
%$\lambda \ge 0$ is the tradeoff parameter to balance the importance between $\mathcal{D}$ and $\mathcal{R}$.  
%posed optimization problem is 
The problem in Eq.~\eqref{eq:optimization} is often solved by first-order  iterative proximal algorithms, \eg fast iterative shrinkage/thresholding algorithm (FISTA)~\cite{beck2009fast} and alternating direction method of multipliers (ADMM) \cite{boyd2011distributed}, to tackle the nonsmoothness of the regularizers.

% Many iterative optimization algorithms can use proximal operators  \cite{beck2009fast,boyd2011distributed, chambolle2011first, parikh2014proximal, geman1995nonlinear, esser2010general}:
To handle the nonsmoothness caused by regularizers, first-order algorithms rely on the proximal operators  \cite{beck2009fast,boyd2011distributed, chambolle2011first, parikh2014proximal, geman1995nonlinear, esser2010general} defined by 
\begin{align}
\mathrm{Prox}_{\sigma^2 \mathcal{R}} (v)  = \mathop{\mathrm{argmin}}_x \big( \mathcal{R}(x) + \frac{1}{2\sigma^2} \| x - v \|_2^2 \big).
\end{align}
Interestingly, given the mathematical equivalence of the proximal operator  to the regularized denoising,  the proximal operators $\mathrm{Prox}_{\sigma^2 \mathcal{R}}$ can be replaced by any off-the-shelf denoisers $\mathcal{H}_\sigma$ with noise level $\sigma$, yielding a new framework namely plug-and-play (PnP) prior \cite{venkatakrishnan2013plug}.  The resulting algorithms, \eg PnP-ADMM, can be written as 
%\begin{align}
%&x^{k+1} = \mathrm{Prox}_{\sigma^2 \mathcal{R}} \left(z^k - u^k \right) = \mathcal{H}_\sigma \left( z^k - u^k  \right) \\
%&z^{k+1} = \mathrm{Prox}_{\frac{1}{\mu} \mathcal{D}} \left(x^{k+1} + u^k \right) \\
%&u^{k+1} = u^k + x^{k+1} - z^{k+1}
%\end{align}
%\begin{equation} \label{eq:pnp-admm}
%  \begin{split}
%&x^{k+1} = \mathrm{Prox}_{\sigma^2 \mathcal{R}} \left(z^k - u^k \right) = \mathcal{H}_\sigma \left( z^k - u^k  \right) \\
%&z^{k+1} = \mathrm{Prox}_{\frac{1}{\mu} \mathcal{D}} \left(x^{k+1} + u^k \right) \\
%&u^{k+1} = u^k + x^{k+1} - z^{k+1}
%  \end{split}
%\end{equation}
\begin{align} 
&x_{k+1} = \mathrm{Prox}_{\sigma_k^2 \mathcal{R}} \left(z_k - u_k \right) = \mathcal{H}_{\sigma_k} \left( z_k - u_k  \right) \label{eq:pnp-admm-1},\\
&z_{k+1} = \mathrm{Prox}_{\frac{1}{\mu_k} \mathcal{D}} \left(x_{k+1} + u_k \right)\label{eq:pnp-admm-2}, \\
&u_{k+1} = u_k + x_{k+1} - z_{k+1}\label{eq:pnp-admm-3},
\end{align}
%where $\sigma = \sqrt{\frac{\lambda}{\mu}}$ and $\mu$ is the penalty parameter to control the strength of the quadratic regularization. 
where $k \in \left[0, \tau \right)$ denotes the $k$-th iteration, $\tau$ is the terminal time, 
%$\sigma_k = \sqrt{\frac{\lambda}{\mu}}$ 
$\sigma_k$ and $\mu_k$ indicate the denoising strength (of the denoiser) and the penalty parameter used in the $k$-th iteration respectively.  

In this formulation, the regularizer $\mathcal{R}$ can be implicitly defined by a plugged denoiser, which opens a new door to leverage the vast progress made on the image denoising front to solve more general inverse imaging problems. To plug well-known image denoisers, \eg BM3D \cite{dabov2007image} and NLM \cite{buades2005non}, into optimization algorithms often leads to sizeable performance gain compared to other explicitly defined regularizers, \eg total variantion. 
That is PnP as a stand-alone framework can combine the  benefits of both deep learning based denoisers and optimization methods,  e.g.~\cite{Zhang_2017_CVPR,Chang_2017_ICCV,Meinhardt_2017_ICCV}. These highly desirable benefits are in terms of \textit{fast and effective inference whilst circumventing the need of expensive network retraining whenever the specific problem changes.}

\begin{figure}[!t]
\centering
\begin{subfigure}[b]{.45\linewidth}
\centering
\includegraphics[width=1\linewidth,clip,keepaspectratio]{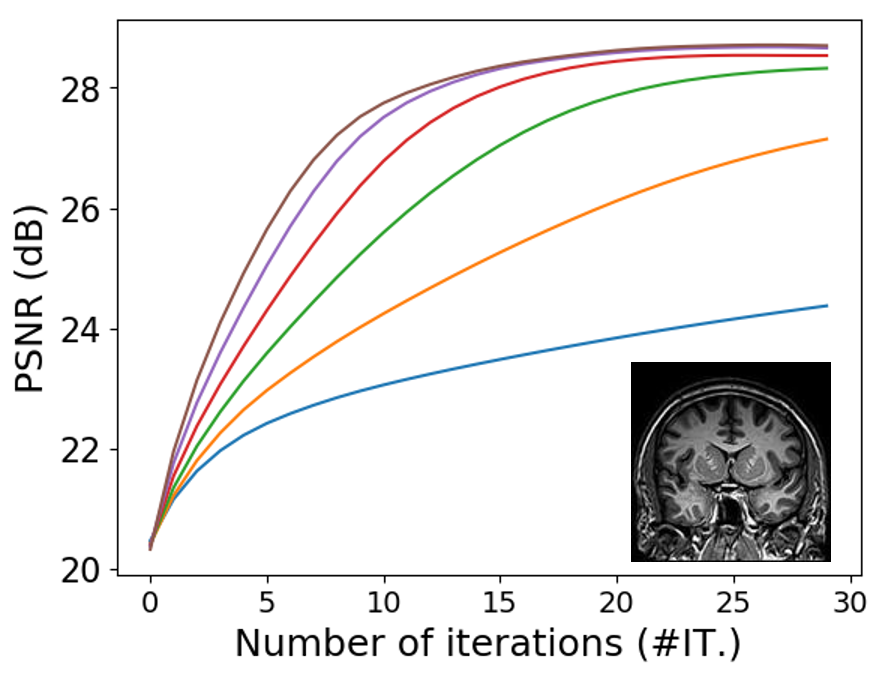}
%\caption{Noisy Input}
\end{subfigure}
\begin{subfigure}[b]{.45\linewidth}
\centering
\includegraphics[width=1\linewidth,clip,keepaspectratio]{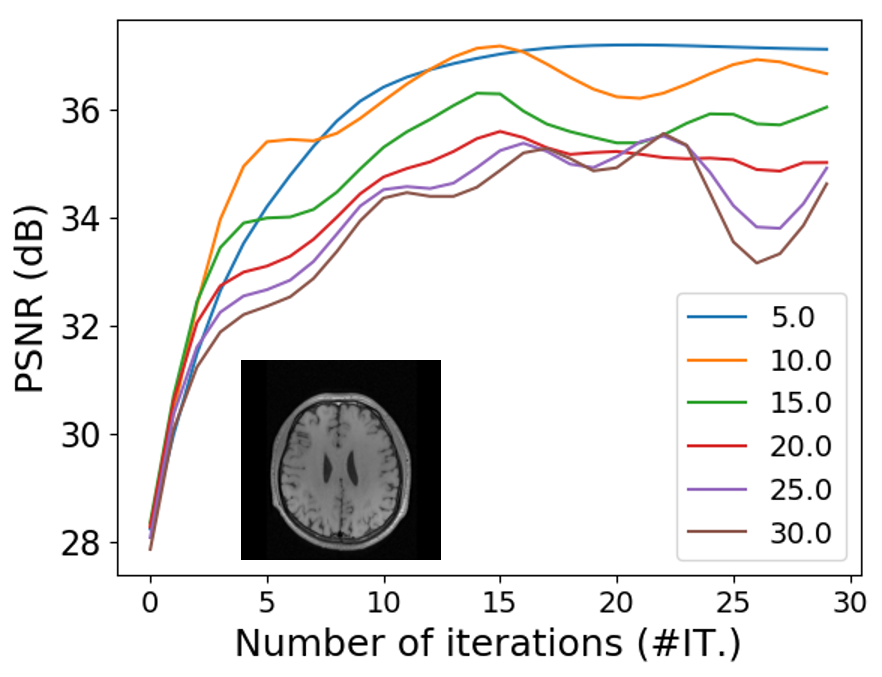}
%\caption{Long-exposure Reference}
\end{subfigure} \\
\begin{subfigure}[b]{.45\linewidth}
\centering
\includegraphics[width=1\linewidth,clip,keepaspectratio]{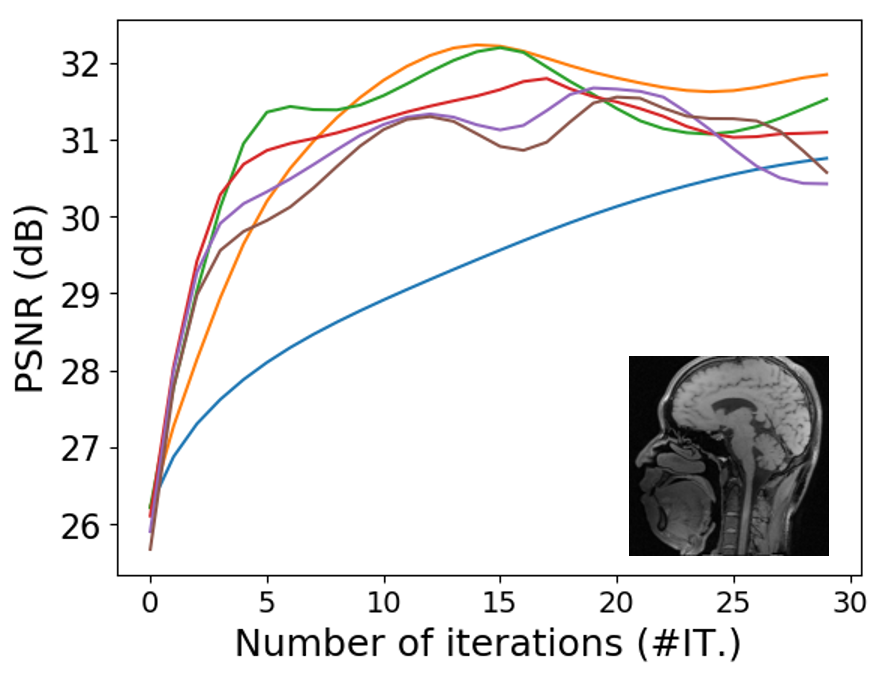}
%\caption{Trained with paired real data}
\end{subfigure}
\begin{subfigure}[b]{.45\linewidth}
\centering
\includegraphics[width=1\linewidth,clip,keepaspectratio]{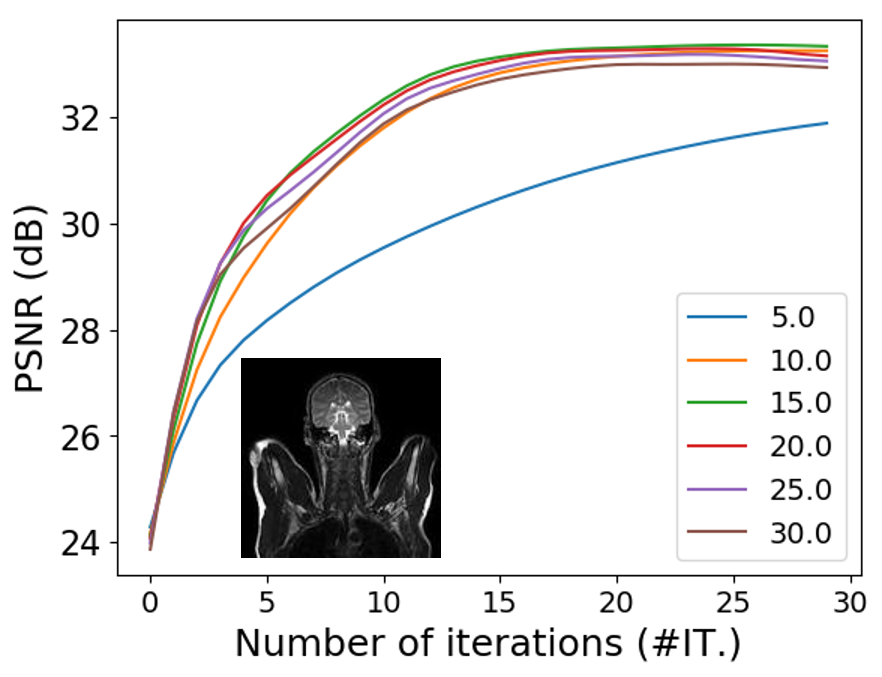}
%\caption{Trained with our synthetic data}
\end{subfigure}
\caption{Compressed Sensing MRI using radial sampling pattern with 20$\%$ sampling rate, where PSNR curves of four medical images are displayed - using PnP-ADMM with different denoising strengths. Different images requires different denoising strengths to reach the optimal performance.} 
\label{fig:example}
\end{figure}

Whilst a PnP framework offers promising image recovery results, a major drawback is that its performance is highly sensitive to the internal parameter selection, which generically includes the penalty parameter $\mu$, 
the denoising strength (of the denoiser) $\sigma$ and the terminal time $\tau$.  The body of literature often utilizes  manual tweaking e.g.~\cite{Chang_2017_ICCV,Meinhardt_2017_ICCV}  or handcrafted criteria e.g.~\cite{chan2017plug,Zhang_2017_CVPR,eksioglu2016decoupled,tirer2018image} to select parameters for each specific problem setting. 
%Previous methods often rely on manual tweaking \cite{Chang_2017_ICCV,Meinhardt_2017_ICCV}  or handcrafted criterion  \cite{chan2017plug,Zhang_2017_CVPR,eksioglu2016decoupled,tirer2018image}  to select parameters for each specific problem setting. 
However, manual parameter tweaking requires several trials, which is very cumbersome and time-consuming. Semi-automated handcrafted criteria (for example monotonically decreasing the denoising strength) can, to some degree, ease the burden of exhaustive search of large parameter space, but often leads to suboptimal local minimum.  
Moreover, the optimal parameter setting differs image-by-image, depending on the measurement model,  noise level,  noise type and  unknown image itself. These differences can be noticed in the further detailed comparison in  Fig.~\ref{fig:example}, where peak signal-to-noise ratio (PSNR) curves are displayed for four images under varying denoising strength.

This paper is devoted to addressing the aforementioned challenge -- how to deal with the manual parameter tuning problem in a PnP framework.  To this end, we formulate the internal parameter selection as a sequential decision-making problem. To do this, a policy is adopted to select a sequence of internal parameters to guide the optimization.
Such problem can be naturally fit into a reinforcement learning (RL) framework, where a policy agent seeks to map observations to actions,  with the aim of maximizing cumulative-reward. The reward reflects the \textit{to do} or \textit{not to do} events for the agent, and a desirable high reward can be obtained if the policy leads to a faster convergence and better restoration accuracy.

\begin{figure}[!t]
\centering
\includegraphics[width=1\linewidth,clip,keepaspectratio]{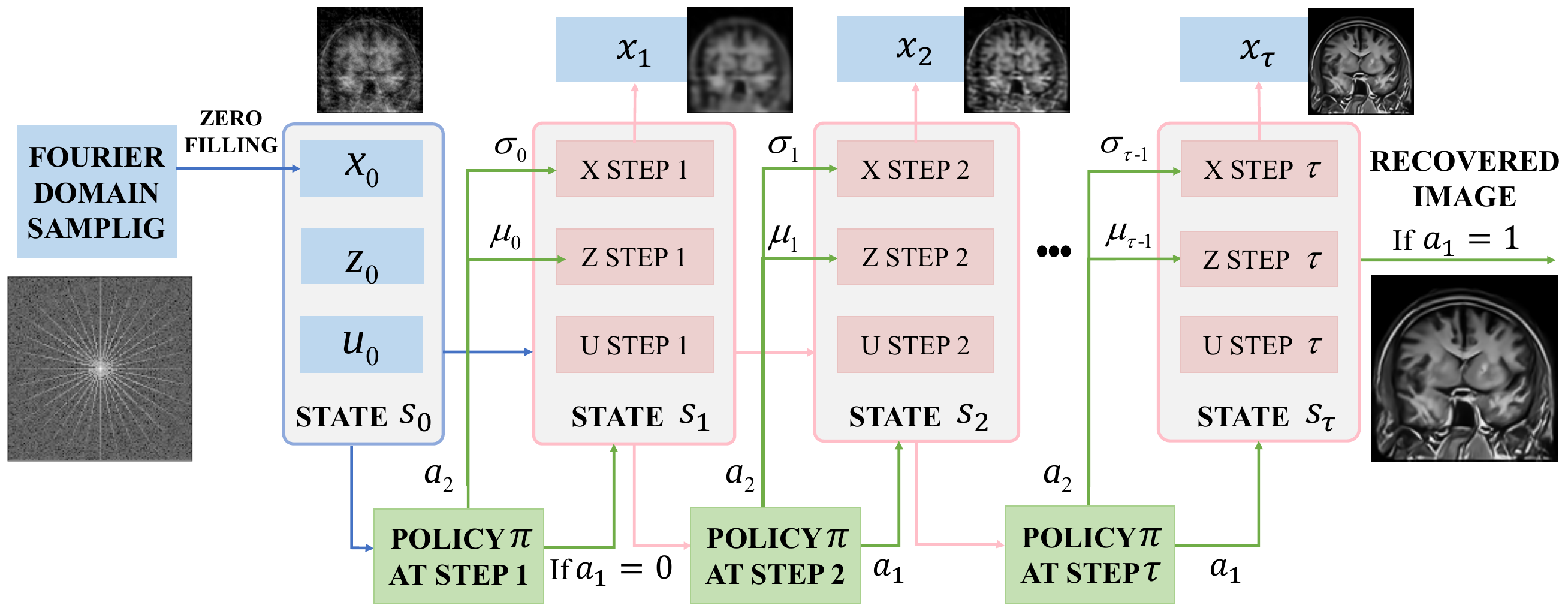}
\caption{Overview of our tuning-free plug-and-play framework (taking CS-MRI problem as example).}
\label{fig:framework}
\end{figure}

We demonstrate, through extensive numerical and visual experiments, the advantage of our algorithmic approach on Compressed Sensing MRI and phase retrieval problems.  We show that the policy well approximates the intrinsic function that maps the input state to its optimal parameter setting. 
By using the learned policy, the guided optimization can reach comparable results to the ones using oracle parameters tuned via the inaccessible ground truth. 
An overview of our algorithm is shown in Fig.~\ref{fig:framework}. 
Our contributions are as follows: 
\begin{packed_enum}
\item We present a tuning-free PnP algorithm that can customize parameters towards diverse images, which often demonstrates faster practical convergence and better empirical performance than handcrafted criteria. 
\item We introduce an efficient mixed model-free and model-based RL algorithm. It can optimize jointly the discrete terminal time, and the continuous denoising strength/penalty parameters.
\item We validate our approach with an extensive range of numerical and visual experiments, and show how the performance of the PnP is affected by the parameters. We also show that our well-designed approach leads to better results than state-of-the-art techniques on compressed sensing MRI and phase retrieval.
%as well as achieving state-of-the-art results on both compressive sensing MRI and phase retrieval. 
\end{packed_enum}

\section{Related Work}
The body of literature has reported several PnP algorithmic techniques. In this section, we provide a short overview of these techniques. 

\noindent {\bf Plug-and-play (PnP).~} 
The definitional concept of PnP was first introduced in  \cite{danielyan2010image,zoran2011learning,venkatakrishnan2013plug}, which has attracted great  attention owing to its effectiveness and flexibility to handle a wide range of inverse imaging problems. Following this philosophy, several works 
%Since then, many follow-up works 
have been developed, and can be roughly categorized in terms of four aspects, i.e., proximal algorithms, imaging applications, denoiser priors, and the convergence. 
\textbf{(i)} \textit{proximal algorithms} include half-quadratic splitting \cite{Zhang_2017_CVPR}, primal-dual method \cite{ono2017primal},  generalized approximate message passing \cite{metzler2016denoising} and (stochastic) accelerated proximal gradient method \cite{sun2019online}.
\textbf{(ii)} \textit{imaging applications} have such as bright field electronic tomography \cite{sreehari2016plug}; diffraction tomography \cite{sun2019online};  low-dose CT imaging \cite{he2018optimizing}; Compressed Sensing MRI \cite{eksioglu2016decoupled};  electron microscopy \cite{sreehari2017multi}; single-photon imaging \cite{chan2017plug}; phase retrieval \cite{metzler2018prdeep}; Fourier ptychography  microscopy \cite{sun2019regularized}; light-field photography \cite{chun2019momentum}; hyperspectral sharpening \cite{teodoro2018convergent}; 
denoising \cite{rond2016poisson};
and image processing  -- e.g. demosaicking, deblurring, super-resolution and inpainting \cite{heide2014flexisp,Meinhardt_2017_ICCV,zhang2019deep,tirer2018image}.  

Moreover, \textbf{(iii)} \textit{denoiser priors} include   BM3D \cite{heide2014flexisp,dar2016postprocessing,rond2016poisson,sreehari2016plug,chan2017plug}, nonlocal means \cite{venkatakrishnan2013plug,heide2014flexisp,sreehari2016plug}, Gaussian mixture models \cite{teodoro2016image,teodoro2018convergent}, weighted nuclear norm minimization \cite{kamilov2017plug}, and deep learning-based denoisers \cite{Meinhardt_2017_ICCV,Zhang_2017_CVPR,Chang_2017_ICCV}. Finally, 
\textbf{(iv)}  \textit{theoretical analysis on the convergence} include  the symmetric gradient  \cite{sreehari2016plug}, the bounded denoiser \cite{chan2017plug} and the nonexpansiveness assumptions \cite{sreehari2016plug,teodoro2018convergent,sun2019online,ryu2019plug,chan2019performance}.

Differing from these aspects, in this work we focus on the challenge of parameter selection in PnP, where a bad choice of parameters often leads to severe degradation of the results \cite{romano2017little,chan2017plug}.
%sometimes even generates unstable or diverged iterates .  
Unlike existing semi-automated parameter tuning criteria \cite{wang2017parameter,chan2017plug,Zhang_2017_CVPR,eksioglu2016decoupled,tirer2018image}, \textit{our method is fully automatic and is purely learned from the data, which significantly eases the burden of manual parameter tuning. }

%The work closely related to us is \cite{tirer2018image}, 
%where an automatic tuning mechanism is present to set parameters. However, their method still requires 
%\noindent {\bf Optimization-inspired deep neural network.~}

\noindent {\bf Automated Parameter Selection.~}
There are some works that considering automatic parameter selection in inverse problems. However, the prior term in these works is restricted to certain
types of regularizers, \eg Tikhonov regularization \cite{hansen1993use,golub1979generalized}, smoothed versions of the $\ell_p$ norm \cite{eldar2008generalized,giryes2011projected}, or general convex functions \cite{ramani2012regularization}.  To the best of our knowledge, none of them can be applicable to the PnP framework with sophisticated non-convex and learned priors.

\noindent {\bf Deep Unrolling.~}
Perhaps the most confusable concept to  PnP in the deep learning era is the so-called deep unrolling methods \cite{gregor2010learning,hershey2014deep,wang2016proximal,NIPS2016_6406,Zhang_2018_CVPR,diamond2017unrolled,NIPS2017_6774,adler2018learned,dong2018denoising,xie2019differentiable}, which explicitly unroll/truncate iterative optimization algorithms into learnable deep architectures. 
In this way, the penalty parameters (and the denoiser prior) are treated as trainable parameters, meanwhile the number of iterations has to be fixed to enable end-to-end training.  \textit{By contrast, our PnP approach can adaptively select a stop time and penalty parameters given varying input states, though using the off-the-shelf denoiser as prior.}

\noindent {\bf Reinforcement Learning for Image Recovery.~}
Although Reinforcement Learning (RL) has been applied in a range of domains, from game playing \cite{mnih2013playing,silver2016mastering} to robotic control \cite{schulman2015trust}, only few works have successfully employed RL to the image recovery tasks. Authors of that~\cite{yu2018crafting} learned a RL policy to select appropriate tools from a toolbox to progressively restore corrupted images. 
The work of~\cite{zhang2018dynamically} proposed a recurrent image restorer whose endpoint was dynamically controlled by a learned policy. In~\cite{furuta2019fully}, authors used RL to select a sequence of classic filters to process images gradually. 
The work of~\cite{yu2019path}  learned network path selection for image restoration in a multi-path CNN.  \textit{In contrast to these works, we apply a mixed model-free and model-based deep RL approach to automatically select the parameters for the PnP image recovery algorithm.}

\section{Tuning-free PnP Proximal Algorithm}
In this work,we elaborate on our tuning-free PnP proximal algorithm, as described in~\eqref{eq:pnp-admm-1}-\eqref{eq:pnp-admm-3}. 
This section describes in detail our approach, which contains three main parts. Firstly, we describe how the automated parameter selection is driven. Secondly, we introduce our environment model, and finally, we introduce the policy learning, which is guided by a mixed model-free and a model-based RL.

It is worth mentioning that our method is generic, and can be applicable to PnP methods derived from other proximal algorithms, \eg forward backward splitting, as well.  The reason is  that these are distinct methods, they share the same fixed points as PnP-ADMM \cite{Meinhardt_2017_ICCV}. 

\subsection{RL Formulation for Automated Parameter Selection} 

This work mainly focuses on the automated parameter selection problem in the PnP framework, where we aim to select a sequence of parameters $(\sigma_0, \mu_0, \sigma_1, \mu_1 ,\cdots,\sigma_{\tau-1}, \mu_{\tau-1}$) to guide optimization such that the recovered image $x^{\tau}$ is close to the underlying image $x$. 
We formulate this problem as a Markov decision process (MDP), which can be addressed via reinforcement learning (RL). 
% Reinforcement learning is a subarea of machine learning related to how an agent should act within an environment, to maximize its cumulative rewards.
%We briefly introduce basic concepts from RL, and how we formulate our automated parameter selection problem as an RL problem. 

We denote the MDP by the tuple $(\mathcal{S}, \mathcal{A}, p, r)$, where $\mathcal{S}$ is the state space, $\mathcal{A}$ is the action space,  $p$ is the transition function describing the environment dynamics, and $r$ is the reward function. 
%Fig.~\ref{fig:framework} illustrates our tuning-free PnP framework taking CS-MRI problem as example. 
Specifically, for our task, $\mathcal{S}$ is the space of optimization variable states, which includes the initialization  $(x_{0}, z_{0}, u_{0})$ and all intermedia results $(x_k, z_k, u_k)$ in the optimization process. $\mathcal{A}$ is the space of internal parameters, including both discrete terminal time $\tau$ and the continuous denoising strength/penalty parameters ($\sigma_k$, $\mu_k$). The transition function $p: \mathcal{S} \times \mathcal{A} \mapsto \mathcal{S}$ maps input state $s \in \mathcal{S}$ to its outcome state $s' \in \mathcal{S}$ after taking action $a \in \mathcal{A}$. The state transition can be expressed as $s_{t+1} = p(s_t, a_t)$, which is composed of one or several iterations of optimization.  On each transition, the environment emits a reward in terms of the reward function $r: S \times \mathcal{A} \mapsto \mathbb{R}$, which evaluates actions given the state. 
Applying a sequence of parameters to the initial state $s_0$ results in a trajectory $T$ of states, actions and rewards: $T = \{s_0, a_0, r_0, \cdots, s_{N}, a_{N}, r_{N}\} $. 
 %where $s_{t} \in \mathcal{S}$, $a_{t} \in \mathcal{A}$, $r_t \in \mathbb{R}$ are states, actions and rewards, $t$ is the time step, $N$ is the number of actions.
 %and $s_{N}$ is the terminal state. 
Given a trajectory $T$, we define the return $r^{\gamma}_{t}$ as the summation of discounted rewards after $s_t$, 
\begin{align}
r^{\gamma}_{t} = \sum_{t'=0}^{N-t} \gamma^{t'} r (s_{t+t'}, a_{t+t'}),
\end{align}
where $\gamma \in [0, 1]$ is a discount factor and prioritizes earlier rewards over later ones. 

Our goal is to learn a policy $\pi$, denoted as $\pi(a|s): \mathcal{S} \mapsto \mathcal{A}$ for the decision-making agent, in order to maximize
the objective defined as 
%  \begin{align}
% J\left(\pi\right) = \mathop{\mathbb{E}}_{\substack{s_0 \sim \mathcal{S}_0\\  T \sim \pi}} \left[ r^{\gamma}_{0} \right]
% \end{align}
 \begin{align}
J\left(\pi\right) = \mathbb{E}_{s_0 \sim S_0,  T \sim \pi} \left[ r^{\gamma}_{0} \right],
\end{align}
where $\mathbb{E}$ represents expectation, $s_0$ is the initial state, and $S_0$ is the corresponding initial state distribution. Intuitively, the objective describes the expected return over all possible trajectories induced by the policy $\pi$. %\cite{sutton2000policy}
The expected return on states and state-action pairs under the policy $\pi$ are defined by state-value functions $V^{\pi}$ and action-value functions $Q^{\pi}$ respectively, i.e.,
% \begin{align}
% V^{\pi}\left(s\right)    &=  \mathop{\mathbb{E}}_{T\sim \pi} \left[ r^{\gamma}_{0} | s_0 = s \right] \\
% Q^{\pi}\left(s, a\right) &= \mathop{\mathbb{E}}_{T\sim \pi} \left[ r^{\gamma}_{0} | s_0 = s, a_0 = a \right]
% \label{eq: Q-function}
% \end{align}
\begin{align}
V^{\pi}\left(s\right)    &=  \mathbb{E}_{T\sim \pi} \left[ r^{\gamma}_{0} | s_0 = s \right], \\
Q^{\pi}\left(s, a\right) &= \mathbb{E}_{T\sim \pi} \left[ r^{\gamma}_{0} | s_0 = s, a_0 = a \right].
\label{eq: Q-function}
\end{align}
%To fit our task into this RL framework, 
In our task, we decompose actions into two parts: a discrete decision $a_1$ on terminal time and a continuous decision $a_2$ on denoising strength and penalty parameter. The policy also consists of two sub-policies: $\pi = (\pi_1, \pi_2)$, a stochastic policy and a deterministic policy that generate $a_1$ and $a_2$ respectively.  
The role of $\pi_1$ is to decide whether to terminate the iterative algorithm when the next state is reached. 
It samples a boolean-valued outcome $a_1$ from a two-class categorical distribution $\pi_1(\cdot | s)$, whose probability mass function is calculated from the current state $s$. 
We move forward to the next iteration if $a_1 = 0$, otherwise the optimization would be terminated to output the final state.  Compared to the stochastic policy $\pi_1$, we treat $\pi_2$ deterministically, \ie $a_2 = \pi_2(s)$ since $\pi_2$ is differentiable with respect to the environment, such that its gradient can be precisely estimated. %

%\subsection{Differentiable Environment} 
\subsection{Environment Model}\label{sec: env-denoiser}

In RL, the environment is characterized by two components: the environment dynamics and  reward function. In our task, the environment dynamics is described by the transition function $p$ related to the PnP-ADMM. Here, we elucidate the detailed setting of the PnP-ADMM as well as the reward function used for training policy. 

%The PnP-ADMM involves three subroutines at each step, i.e., a denoising step in ~\eqref{eq:pnp-admm-1} to enforce image prior, a proximal mapping to enforce data-consistency in ~\eqref{eq:pnp-admm-2}, and an update of the scaled dual variable in ~\eqref{eq:pnp-admm-3}. 

\noindent {\bf Denoiser Prior.~} 
Differentiable environment makes the policy learning more efficient. 
To make the environment differentiable with respect to $\pi_2$\footnote{$\pi_1$ is non-differentiable towards environment regardless of the formulation of the environment.}, 
we take a convolutional neural network (CNN) denoiser as the image prior. 
In practice, we use a residual U-Net~\cite{ronneberger2015u}  architecture, which was originally designed for medical image segmentation, but was founded to be useful in image denoising recently.  Besides, 
we incorporate an additional tunable noise level map into the input as  \cite{zhang2018ffdnet}, enabling us to provide continuous noise level control (\ie different denoising strength) within a single network. 

\noindent {\bf Proximal operator of data-fidelity term.~} 
Enforcing consistency with measured data requires evaluating the proximal operator in ~\eqref{eq:pnp-admm-2}. For  inverse problems, there might exist fast solutions due to the special structure of the observation model. We adopt the fast solution if feasible (\eg closed-form solution using fast Fourier transform, rather than the general matrix inversion) 
otherwise a single step of gradient descent is performed as an inexact solution for~\eqref{eq:pnp-admm-2}.
%We use a single step of gradient descent as an inexact solution for the otherwise computationally expensive general case. 
%the general solution involves a large-scale matrix inversion, which is computationally expensive. 

\noindent {\bf Transition function.~} To reduce the computation cost, we define the transition function $p$ to involve $m$ iterations of the optimization. 
At each time step, the agent thus needs to decide the internal parameters for $m$ iterates.
We set $m=5$ and the max time step $N=6$ in our algorithm, leading to 30 iterations of the optimization at most. 

\noindent {\bf Reward function.~}
To take both image recovery performance and runtime efficiency into account,
we define the reward function as 
\begin{align}
r(s_t, a_t) = \zeta(p(s_t, a_t)) - \zeta(s_t) - \eta. %(1 - I_{\{1\}}(a_t))
\end{align}
The first term, $\zeta(p(s_t, a_t)) - \zeta(s_t)$,  
denotes the PSNR increment made by the policy, where $\zeta(s_t)$ denotes the PSNR of the recovered image at step $t$. 
A higher reward is acquired if the policy leads to higher performance gain in terms of PSNR. 
The second term, $\eta$, implies penalizing the policy as it does not select to terminate at step $t$, where $\eta$ sets the degree of penalty.
A negative reward is given if the PSNR gain does not exceed the degree of penalty, thereby encouraging the policy to early stop the iteration with diminished return.  We set $\eta=0.05$ in our algorithm\footnote{The choice of the hyperparameters  $m,N$ and $\eta$ is discussed in the \textit{suppl. material}.}.

%\begin{algorithm}[!t]
 % \caption{Training procedure} 
  %\label{alg:training}
  %\small
  %\begin{algorithmic}[1] 
  %\REQUIRE 
  %initial state distribution $S$, state buffer $D$, initialized network parameters $\theta$, $\phi$, $\hat{\phi}$, learning rates $lr_{\theta}$, $lr_{\phi}$. $\beta$ for updating target value network.
   %\FOR{each iteration}
    %\STATE sample initial state $s_0$ from $S$
     %\FOR{environment step $t \in [1, N]$}
  	   % \STATE $a_t \sim \pi_{\theta} (a_t | s_t) $
  	    %\STATE $s_{t+1} \sim p(s_{t+1} | s_t, a_t) $
  	    %\STATE $D \leftarrow D \cup \{s_{t+1}\}$
     %\ENDFOR
     %\FOR{each gradient step}
      %  \STATE sample states from the state buffer $D$
      %  \STATE $\theta_1 \leftarrow \theta_1 + lr_{\theta} \triangledown_{\theta_1} J(\pi_\theta)  $
      %  \STATE $\theta_2 \leftarrow \theta_2 + lr_{\theta} \triangledown_{\theta_2} J(\pi_\theta)  $
      %  \STATE $\phi \leftarrow \phi - lr_{\phi} \triangledown_\theta L_{\phi} $
      %  \STATE $\hat{\phi} \leftarrow \beta \phi + (1-\beta) \hat{\phi}$
     %\ENDFOR
    %\ENDFOR
  %\end{algorithmic}
%\end{algorithm}

\subsection{RL-based policy learning}
In this section, we present a mixed model-free and model-based RL algorithm to learn the policy.
Specifically, model-free RL (agnostic to the environment dynamics) is used to train $\pi_1$, while model-based RL is utilized to optimize $\pi_2$ to make full use of the environment model\footnote{$\pi_2$ can also be optimized in a model-free manner. The comparison can be found in the Section \ref{sec:csmri}.}. 
We apply the actor-critic framework \cite{sutton2000policy}, 
that uses a policy network $\pi_\theta(a_t| s_t)$ (actor) and a value network $V_\phi^{\pi}(s_t)$ (critic) to formulate the policy and the state-value function respectively.  
For convenience, we follow \cite{Huang_2019_ICCV} that uses residual structures similar to ResNet-18 \cite{He_2016_CVPR} as the feature extractor in the policy
and value networks, followed by fully-connected layers and activation functions to produce desired outputs\footnote{Details of networks are given in the \textit{suppl. material}.}.
%that formulates both the policy and the state-value function by CNN-based function approximators\footnote{Details of networks are given in the \textit{suppl. material}.}, namely actor and critic respectively.  We use a policy network $\pi_\theta(a_t| s_t)$ and a value network $V_\phi^{\pi}(s_t)$, parameterized by $\theta$ and $\phi$ respectively.
The policy and the value networks are learned in an interleaved manner.  
For each gradient step, we optimize the value network parameters $\phi$ by minimizing
% \begin{align}
%     L_{\phi} = \mathbb{E}_{s \sim \rho^\pi, a \sim \pi(s)} \left[ \frac{1}{2} (r(s, a) + \gamma V^{\pi}_{\hat{\phi}}(p(s, a)))^2 \right]
% \end{align}
% where $\rho^\pi$ is the discounted state distribution defined as: $\rho^\pi = \sum_{k=0}^{\infty} \gamma^k\mathbb{P} (s_k = s) $.
\begin{align}
    L_{\phi} = \mathbb{E}_{s \sim B, a \sim \pi_{\theta}(s)} \left[ \frac{1}{2} (r(s, a) + \gamma V^{\pi}_{\hat{\phi}}(p(s, a)) - V^{\pi}_{\phi}(s) )^2 \right],
\end{align}
where $B$ is the distribution of previously sampled states, practically implemented by a state buffer. This partly serves as a role of the experience replay mechanism \cite{lin1992self-improving}, which is observed to "smooth" the training data distribution \cite{mnih2013playing}. The update makes use of a target value network $V^{\pi}_{\hat{\phi}}$, where $\hat{\phi}$ is the exponentially moving average of the value network weights and has been shown to stabilize training \cite{mnih2015human-level}. 

The policy network has two sub-policies, which employs shared convolutional layers to extract image features, followed by two separated groups of fully-connected layers to produce termination probability $\pi_1(\cdot|s)$ (after softmax) or denoising strength/penalty parameters $\pi_2(s)$ (after sigmoid). We denote the parameters of the sub-polices as $\theta_1$ and $\theta_2$ respectively, and we seek to optimize $\theta=(\theta_1, \theta_2)$ so that the objective $J(\pi_\theta)$ is maximized. 
The policy network is trained using policy gradient methods \cite{peters2006policy}. 
%which maximize the expected return by following the gradient of this expectation with respect to the policy parameters. 
The gradient of $\theta_1$ is estimated by a likelihood estimator in a model-free manner, while the gradient of $\theta_2$ is estimated relying on backpropagation via environment dynamics in a model-based manner. 
Specifically, for discrete terminal time decision $\pi_1$, 
we apply the policy gradient theorem \cite{sutton2000policy} to obtain unbiased Monte Carlo estimate of $\triangledown_{\theta_1} J(\pi_\theta)$ using advantage function $A^\pi(s, a) = Q^\pi(s, a) - V^\pi(s)$ as target, i.e.,  
% \begin{align}
% \triangledown_{\theta_1} J(\pi_\theta)= &\mathop{\mathbb{E}}_{\substack{s \sim \rho^\pi \\ a_1 \sim \pi_1(s) \\ a_2 = \pi_2(s)}}
% \left[\triangledown_{\theta_1} \mathrm{log}\,\pi_1(a_1|s)\,A^{\pi}(s, (a_1, a_2))\right]
% \end{align}
\begin{align}
\triangledown_{\theta_1} J(\pi_\theta)= &\mathbb{E}_{s \sim B, a \sim \pi_{\theta}(s)}
\left[\triangledown_{\theta_1} \mathrm{log}\,\pi_1(a_1|s)\,A^{\pi}(s, a)\right]. 
\label{eq:pi_1}
\end{align}
  For continuous denoising strength and penalty parameter selection $\pi_2$,
we utilize the deterministic policy gradient theorem \cite{silver2014deterministic} to formulate its gradient, i.e.,  
% \begin{align}
% \triangledown_{\theta_2} J(\pi_\theta)= &\mathop{\mathbb{E}}_{\substack{s \sim \rho^\pi \\ a_1 \sim \pi_1(s) \\ a_2 = \pi_2(s)}}
% \left[\triangledown_{a_2} \,Q^{\pi}(s, a)\triangledown_{\theta_2} \pi_2 (s) \right]
% \end{align}
\begin{align}
\triangledown_{\theta_2} J(\pi_\theta)= &\mathbb{E}_{s \sim B, a \sim \pi_{\theta}(s)}
\left[\triangledown_{a_2} \,Q^{\pi}(s, a)\triangledown_{\theta_2} \pi_2 (s) \right],
\label{eq:pi_2}
\end{align}
where we approximate the action-value function $Q^{\pi}(s, a)$ by $r(s, a) + \gamma V_\phi^{\pi}(p(s, a))$ given its unfolded definition. 

Using the chain rule, we can directly obtain the gradient of $\theta_2$ by backpropagation via the reward function, the value network and the transition function, in contrast to relying on the gradient backpropagated from only the learned action-value function in the model-free DDPG algorithm \cite{lillicrap2015continuous}. 
%The whole training procedure is summarized in Alg. \ref{alg:training}.
% This methods alternates between collecting experience from the environment with the current policy, and updating both the policy and the value networks using the gradients from batches sampled from a state buffer. 
% In practice, we take a single environment step followed by ten gradient steps. 

\begin{table}[t]
    \centering
    \caption{Comparisons of different CNN-based denoisers: we show the results of (1) Gaussian denoising performance (PSNR) under noise level $\sigma=50$; (2) the CS-MRI performance (PSNR) when plugged into the PnP-ADMM; (3) the GPU runtime (ms) of denoisers when processing an image with size $256\times256$.}
\begin{tabular}{c|c|c|c|}
%\cline{2-4}
\hline
\multicolumn{1}{|c|}{Performance}                 & \cellcolor[HTML]{EFEFEF}DnCNN & \cellcolor[HTML]{EFEFEF}MemNet & \cellcolor[HTML]{EFEFEF}UNet \\ \hline
\multicolumn{1}{|c|}{\textsc{Denoising Perf.}} & 27.18                         & 27.32                          & 27.40                       \\ \hline
\multicolumn{1}{|c|}{\textsc{PnP Perf.}}       & 25.43                         & 25.67                          & 25.76                       \\ \hline
\multicolumn{1}{|c|}{\textsc{Times}}           & 8.09                          & 64.65                          & 5.65                      \\ \hline
\end{tabular}
    \label{tb:denoiser-profile}
\end{table}

\begin{table}[t]
\centering
\caption{Comparisons of different policies used in PnP-ADMM algorithm for CS-MRI on seven widely used medical images under various acceleration factors (x2/x4/x8) and noise level 15. 
We show both PSNR and the number of iterations (\#IT.) used to induce results. * denotes to report the best PSNR over all iterations (\ie with optimal early stopping). 
The best results are indicated by \textcolor{orange}{orange} color and the second best results are denoted by \textcolor{blue}{blue} color. 
}
\footnotesize
\resizebox{0.45\textwidth}{!}{
\begin{tabular}{lcccccc} 
		\toprule
		 & \multicolumn{2}{c}{$\times2$} & \multicolumn{2}{c}{$\times4$} & \multicolumn{2}{c}{$\times8$} \\ \cline{2-7} 
		\textsc{Policies} & PSNR & \#IT.  & PSNR & \#IT.  & PSNR & \#IT. \\  \midrule
		handcrafted & 30.05 & 30.0 & 27.90 & 30.0 & 25.76 & 30.0\\
		handcrafted$^{*}$ & 30.06 & 29.1 & 28.20 & 18.4 & 26.06 & 19.4\\
		fixed  & 23.94 & 30.0 & 24.26 & 30.0 & 22.78 & 30.0 \\
		fixed$^{*}$  & 28.45 & 1.6 & 26.67 & 3.4 & 24.19 & 7.3 \\
		fixed optimal & 30.02 & 30.0 & 28.27 & 30.0 & 26.08 & 16.7 \\
		fixed optimal$^{*}$ & 30.03 & 6.7 & 28.34 & 12.6 & 26.16 & 30.0 \\
		oracle & 30.25 & 30.0 & 28.60 & 30.0 & 26.41 & 30.0 \\
		oracle$^{*}$ & \textcolor{blue}{30.26} & 8.0 & \color{orange}{28.61} & 13.9 & \color{orange}{26.45} & 21.6  \\ \midrule
		model-free & 28.79 & 30.0 & 27.95 & 30.0 & 26.15 & 30.0 \\
	    Ours  & \color{orange}{30.33} & 5.0 & \textcolor{blue}{28.42} & 5.0 & \color{blue}{26.44} & 15.0 \\
		\bottomrule
\end{tabular}}
\label{tb:policy-comparison}
\end{table}

\begin{table*}[t]
	\centering
	\caption{Quantitative results (PSNR) of different CS-MRI methods on two datasets under various acceleration factors $f$ and noise levels $\sigma_n$. 
	The best results are indicated by \textcolor{orange}{orange} color and the second best results are denoted by \textcolor{blue}{blue} color.  
	}
	\footnotesize
	\setlength{\tabcolsep}{1mm}{
	\begin{tabular}{|C{.095\linewidth}|C{.05\linewidth}|C{.05\linewidth}|C{.095\linewidth}C{.095\linewidth}C{.105\linewidth}C{.105\linewidth}C{.115\linewidth}C{.085\linewidth}C{.085\linewidth}|}
		\hline
		\multirow{2}{*}{\textsc{Dataset}} & \multirow{2}{*}{$f$} & \multirow{2}{*}{$\sigma_n$}  & \multicolumn{2}{c|}{ \cellcolor[HTML]{EFEFEF}\textsc{Traditional}} & \multicolumn{2}{c|}{ \cellcolor[HTML]{EFEFEF}\textsc{Deep Unrolling}} & \multicolumn{3}{c|}{ \cellcolor[HTML]{EFEFEF}\textsc{PnP}} \\ \cline{4-10}
		& & &  RecPF & FCSA & ADMMNet  & ISTANet & BM3D-MRI & IRCNN & Ours \\\hline 
\multirow{9}{*}{Medical7}  
&\multirow{3}{*}{$\times 2$} &$5$ &$32.46$ &$31.70$ &$33.10$ &$34.58$ &$33.33$ &\textcolor{blue}{$34.67$} &\textcolor{orange}{$34.78$}\\%\cline{3-10}
& &$10$ &$29.48$ &$28.33$ &$31.37$ &\textcolor{blue}{$31.81$} &$29.44$ &$31.80$ &\textcolor{orange}{$32.00$}\\%\cline{3-10}
& &$15$ &$27.08$ &$25.52$ &$29.16$ &\textcolor{blue}{$29.99$} &$26.90$ &$29.96$ &\textcolor{orange}{$30.27$}\\\cline{2-10}
&\multirow{3}{*}{$\times 4$} &$5$ &$28.67$ &$28.21$ &$30.24$ &$31.34$ &$30.33$ &\textcolor{blue}{$31.36$} &\textcolor{orange}{$31.62$}\\%\cline{3-10}
& &$10$ &$26.98$ &$26.67$ &$29.20$ &\textcolor{orange}{$29.71$} &$28.30$ &$29.52$ &\textcolor{blue}{$29.68$}\\%\cline{3-10}
& &$15$ &$25.58$ &$24.93$ &$27.87$ &\textcolor{blue}{$28.38$} &$26.66$ &$27.94$ &\textcolor{orange}{$28.43$}\\\cline{2-10}
&\multirow{3}{*}{$\times 8$} &$5$ &$24.72$ &$24.62$ &$26.57$ &\textcolor{blue}{$27.65$} &$26.53$ &$27.32$ &\textcolor{orange}{$28.26$}\\%\cline{3-10}
& &$10$ &$23.94$ &$24.04$ &$26.21$ &\textcolor{blue}{$26.90$} &$25.81$ &$26.44$ &\textcolor{orange}{$27.35$}\\%\cline{3-10}
& &$15$ &$23.18$ &$23.36$ &$25.49$ &\textcolor{blue}{$26.23$} &$25.09$ &$25.53$ &\textcolor{orange}{$26.41$}\\\hline
\multirow{9}{*}{MICCAI}  
&\multirow{3}{*}{$\times 2$} &$5$ &$36.39$ &$34.90$ &$36.74$ &$38.17$ &$36.00$ &\textcolor{blue}{$38.42$} &\textcolor{orange}{$38.57$}\\%\cline{3-10}
& &$10$ &$31.95$ &$30.12$ &$34.20$ &$34.81$ &$31.39$ &\textcolor{blue}{$34.93$} &\textcolor{orange}{$35.06$}\\%\cline{3-10}
& &$15$ &$28.91$ &$26.68$ &$31.42$ &$32.65$ &$28.46$ &\textcolor{blue}{$32.81$} &\textcolor{orange}{$33.09$}\\\cline{2-10}
&\multirow{3}{*}{$\times 4$} &$5$ &$33.05$ &$32.30$ &$34.15$ &$35.46$ &$34.79$ &\textcolor{blue}{$35.80$} &\textcolor{orange}{$36.11$}\\%\cline{3-10}
& &$10$ &$30.21$ &$29.56$ &$32.58$ &\textcolor{orange}{$33.13$} &$31.63$ &$32.99$ &\textcolor{blue}{$33.07$}\\%\cline{3-10}
& &$15$ &$28.13$ &$26.93$ &$30.55$ &\textcolor{orange}{$31.48$} &$29.35$ &$30.98$ &\textcolor{blue}{$31.42$}\\\cline{2-10}
&\multirow{3}{*}{$\times 8$} &$5$ &$28.35$ &$28.71$ &$30.36$ &$31.62$ &$31.34$ &\textcolor{blue}{$31.66$} &\textcolor{orange}{$32.64$}\\%\cline{3-10}
& &$10$ &$26.86$ &$27.68$ &$29.78$ &\textcolor{blue}{$30.54$} &$29.86$ &$30.16$ &\textcolor{orange}{$30.89$}\\%\cline{3-10}
& &$15$ &$25.70$ &$26.35$ &$28.83$ &\textcolor{blue}{$29.50$} &$28.53$ &$28.72$ &\textcolor{orange}{$29.65$}\\\hline
	\end{tabular}}
	\label{tb:CS-MRI}
\end{table*}

\section{Experiments}
In this section, we detail the experiments and evaluate our proposed algorithm. We mainly focus on the tasks of Compressed Sensing MRI (CS-MRI) and phase retrieval (PR), which are the representative linear and nonlinear inverse imaging problems respectively. 

\subsection{Implementation Details}
Our algorithm requires two training processes for: the denoising network and the policy network (and value network). 
For training the denoising network, we follow the common practice that uses 87,000 overlapping patches (with size $128\times128$) drawn from 400 images from the BSD dataset \cite{MartinFTM01}. For each patch, we add white Gaussian noise with noise level sampled from $[1, 50]$. 
The denoising networks are trained with 50 epoch using $L_1$ loss and Adam optimizer \cite{kingma2014adam} with batch size 32. The base learning rate is set to $10^{-4}$ and halved at epoch 30, then reduced to $10^{-5}$ at epoch 40.

To train the policy network and value network, we use the 17,125 resized images with size $128\times128$ from the PASCAL VOC dataset \cite{everingham2014pascal}. 
Both networks are trained using Adam optimizer with batch size 48 and 1500 iterations, with a base learning rate of $3\times10^{-4}$ for the policy network and $10^{-3}$ for the value network. Then we set these learning rates to $10^{-4}$ and $3\times10^{-4}$ at iteration 1000.
We perform 10 gradient steps at every iteration.
%\footnote{More details are available in the \textit{suppl. material.}}

For the  CS-MRI application, a single policy network is trained to handle multiple sampling ratios (with x2/x4/x8 acceleration) and noise levels (5/10/15), simultaneously. Similarly, one policy network is learned for phase retrieval under different settings.

\begin{figure*}[t]
	\centering
	\setlength\tabcolsep{1.5pt}
	\begin{tabular}{cccccccc}		
	 RecPF & FCSA & ADMMNet & ISTANet & BM3D-MRI & IRCNN  & Ours & GroundTruth  \\
	\includegraphics[width=0.12\linewidth]{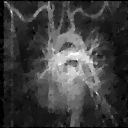}
	& \includegraphics[width=0.12\linewidth]{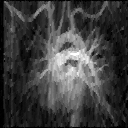}
	& \includegraphics[width=0.12\linewidth]{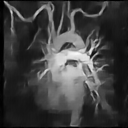}
	& \includegraphics[width=0.12\linewidth]{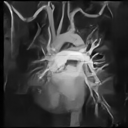}
	& \includegraphics[width=0.12\linewidth]{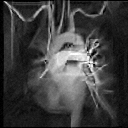}
	& \includegraphics[width=0.12\linewidth]{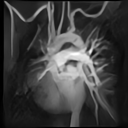}
	& \includegraphics[width=0.12\linewidth]{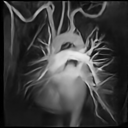}
	& \includegraphics[width=0.12\linewidth]{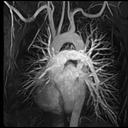} \\
	22.57 & 22.27 & 24.15 & 24.61 & 23.64 & 24.16 & 25.28 & PSNR \\
	\includegraphics[width=0.12\linewidth]{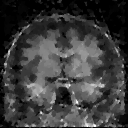}
	& \includegraphics[width=0.12\linewidth]{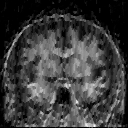}
	& \includegraphics[width=0.12\linewidth]{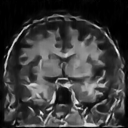}
	& \includegraphics[width=0.12\linewidth]{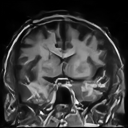}
	& \includegraphics[width=0.12\linewidth]{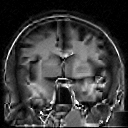}
	& \includegraphics[width=0.12\linewidth]{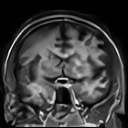}
	& \includegraphics[width=0.12\linewidth]{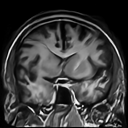}
	& \includegraphics[width=0.12\linewidth]{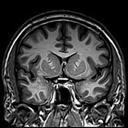} \\
	18.74 & 19.23 & 20.48 & 21.37 & 20.62 & 20.91 & 22.02 & PSNR \\
	\includegraphics[width=0.12\linewidth]{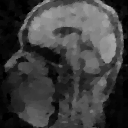}
	& \includegraphics[width=0.12\linewidth]{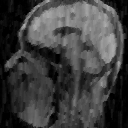}
	& \includegraphics[width=0.12\linewidth]{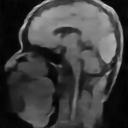}
	& \includegraphics[width=0.12\linewidth]{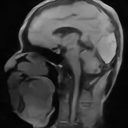}
	& \includegraphics[width=0.12\linewidth]{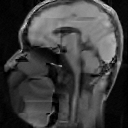}
	& \includegraphics[width=0.12\linewidth]{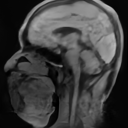}
	& \includegraphics[width=0.12\linewidth]{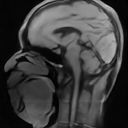}
	& \includegraphics[width=0.12\linewidth]{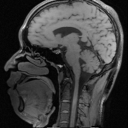} \\
	24.89 & 24.47 & 26.85 & 27.90 & 26.72 & 27.74 & 28.65 & PSNR \\
	\end{tabular} 
	\caption{CS-MRI reconstruction results of different algorithms on medical images.  (\textbf{best view on screen with zoom}). }
	\label{fig:comparison-cs-mri}
\end{figure*}

\subsection{Compressed sensing MRI} \label{sec:csmri}
The forward model of CS-MRI can be mathematically described as $ y = \mathcal{F}_p x + \omega$, 
% \begin{align}
% y = \mathcal{F}_p x + \omega
% \end{align}
where $x \in \mathbb{C}^N$ is the underlying image,  the operator $\mathcal{F}_p : \mathbb{C}^N \rightarrow \mathbb{C}^M$, with $M < N$, denotes the partially-sampled Fourier transform, and $\omega \sim \mathcal{N} \left(0, \sigma_n I_M \right)$ is the additive white Gaussian noise. The data-fidelity term is $\mathcal{D} (x) = \frac{1}{2} \| y - \mathcal{F}_p x \|^2$
%  \begin{align}
% \mathcal{D} (x) = \frac{1}{2} \| y - \mathcal{F}_p x \|^2
% \end{align}
whose proximal operator is given in \cite{eksioglu2016decoupled}.

%\noindent {\bf Practical considerations on denoiser priors.~} 
\noindent {\bf Denoiser priors.~} 
To show how denoiser priors affect the performance of the PnP, we train three state-of-the-art CNN-based denoisers, \ie DnCNN \cite{zhang2017beyond}, MemNet \cite{Tai_2017_ICCV} and residual UNet \cite{ronneberger2015u}, with tunable noise level map. We compare both the Gaussian denoising performance and the PnP performance\footnote{We exhaustively search the best denoising strength/penalty parameters to exclude the impact of internal parameters.} using these denoisers. 
As shown in Table \ref{tb:denoiser-profile}, the residual UNet and MemNet consistently outperform DnCNN in terms of denoising and CS-MRI. It seems to imply a better Gaussian denoiser is also a better denoiser prior for the PnP framework\footnote{Further investigation of this argument can be found in the \textit{suppl. material}.}. Since UNet is significantly faster than MemNet, we choose UNet as our denoiser prior. 
% We also investigate two training strategies used in prior works, \ie training with (1) (real) spectral normalization \cite{ryu2019plug} and (2) adversarial GAN loss \cite{Chang_2017_ICCV}.
% As the internal parameters would be properly selected in our algorithm, we find that there is no need to use (real) spectral normalization \cite{ryu2019plug} to enforce Lipschitz condition in training denoising networks. We also find that  training with adversarial GAN loss as \cite{Chang_2017_ICCV} would adversely impact the PnP performance.  

\begin{figure*}[!htbp]
	\centering
	\setlength\tabcolsep{1.5pt}
	\begin{tabular}{cccccccc}		
	HIO & WF & DOLPHIn & SPAR & \footnotesize{BM3D-prGAMP} & prDeep & Ours & GroundTruth   \\
% 	\includegraphics[width=0.12\linewidth]{figures/PR/Ecoli/HIO.png}
% 	& \includegraphics[width=0.12\linewidth]{figures/PR/Ecoli/WF.png}
% 	& \includegraphics[width=0.12\linewidth]{figures/PR/Ecoli/DOLPHIn.png}
% 	& \includegraphics[width=0.12\linewidth]{figures/PR/Ecoli/SPAR.png}
% 	& \includegraphics[width=0.12\linewidth]{figures/PR/Ecoli/BM3D-prGAMP.png}
% 	& \includegraphics[width=0.12\linewidth]{figures/PR/Ecoli/prDeep.png}
% 	& \includegraphics[width=0.12\linewidth]{figures/PR/Ecoli/Ours.png}
% 	& \includegraphics[width=0.12\linewidth]{figures/PR/Ecoli/gt.png} \\
	\includegraphics[width=0.12\linewidth]{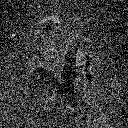}
	& \includegraphics[width=0.12\linewidth]{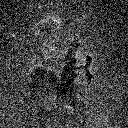}
	& \includegraphics[width=0.12\linewidth]{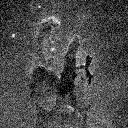}
	& \includegraphics[width=0.12\linewidth]{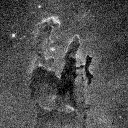}
	& \includegraphics[width=0.12\linewidth]{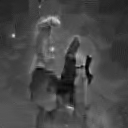}
	& \includegraphics[width=0.12\linewidth]{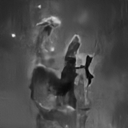}
	& \includegraphics[width=0.12\linewidth]{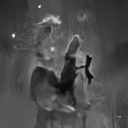}
	& \includegraphics[width=0.12\linewidth]{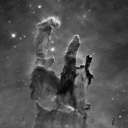} \\
    14.40 & 15.52 & 19.35 & 22.48 & 25.66 & 27.72 & 28.01 & PSNR \\
	\includegraphics[width=0.12\linewidth]{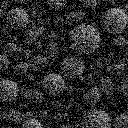}
	& \includegraphics[width=0.12\linewidth]{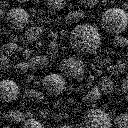}
	& \includegraphics[width=0.12\linewidth]{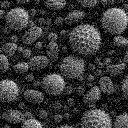}
	& \includegraphics[width=0.12\linewidth]{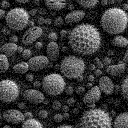}
	& \includegraphics[width=0.12\linewidth]{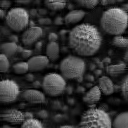}
	& \includegraphics[width=0.12\linewidth]{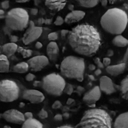}
	& \includegraphics[width=0.12\linewidth]{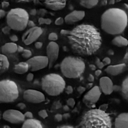}
	& \includegraphics[width=0.12\linewidth]{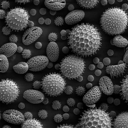} \\
	15.10 & 16.27 & 19.62 & 22.51 & 23.61 & 24.59 & 25.12 & PSNR \\
	\end{tabular} 
	\caption{Recovered images from noisy intensity-only CDP measurements with seven PR algorithms. (\textbf{Details are better appreciated on screen.}). }
	\label{fig:comparison-phase-retrieval}
\end{figure*}

\begin{table}[!htbp]
\centering
\caption{Quantitative results of different PR algorithms on four CDP measurements and varying amount of Possion noise (large $\alpha$ indicates low sigma-to-noise ratio).}
\footnotesize
\begin{tabular}{lccc} 
		\toprule
		 & $\alpha=9$ & $\alpha=27$ & $\alpha=81$ \\ \cline{2-4} 
		Algorithms & PSNR  & PSNR  & PSNR  \\  \midrule
		HIO & 35.96 & 25.76 & 14.82 \\
		WF  & 34.46 &  24.96 & 15.76 \\
		DOLPHIn & 29.93 & 27.45 & 19.35 \\
		SPAR & 35.20 & 31.82 & 22.44 \\
		BM3D-prGAMP & \textcolor{blue}{40.25} & 32.84 & 25.43 \\
		prDeep  & 39.70 & \textcolor{blue}{33.54} & \textcolor{blue}{26.82} \\
		Ours  & \textcolor{orange}{40.33} & \textcolor{orange}{33.90} & \textcolor{orange}{27.23} \\ 
		\bottomrule
\end{tabular}
\label{tb:phase-retrieval}
\end{table}

\noindent {\bf Comparisons of different policies.~} 
We start by giving some insights of our learned policy  by comparing the performance
%To better evaluate the performance of the learned policy, we compare the performance 
of PnP-ADMM with different polices: i) the handcrafted policy
% \footnote{We manually select the best hyperparameters for this semi-automated criteria.}
used in IRCNN~\cite{Zhang_2017_CVPR}; ii) the fixed policy that uses fixed parameters ($\sigma=15$, $\mu=0.1$); iii) the fixed optimal policy that adopts fixed parameters searched to maximize the average PSNR across all testing images; iv) the oracle policy that uses different parameters for different images such that the PSNR of each image is maximized and  v) our learned policy based on a learned policy network to optimize parameters for each image. We remark that all compared polices are run for 30 iteration whilst ours automatically choose the terminal time.
%Except our policy that can automatically choose terminal time,  other polices are all run for 30 iterations. 
% 

To understand the usefulness of the early stopping
mechanism, we also report the results of these polices with optimal early stopping\footnote{It should be noted some policies (\eg "fixed optimal"
and "oracle") requires to access the ground truth to determine parameters, which is generally impractical in real testing scenarios. }.
Moreover, we analyze whether the model-based RL benefits our algorithm by comparing it with the learned policy by model-free RL whose $\pi_2$ is optimized using the model-free DDPG algorithm \cite{lillicrap2015continuous}.
%Besides, to analyze whether the model-based RL benefits our algorithm. We compare our algorithm with the learned policy by model-free RL whose $\pi_2$ is optimized using the model-free DDPG algorithm \cite{lillicrap2015continuous}.

The results of all aforementioned policies are provided in 
Table \ref{tb:policy-comparison}. We can see that the bad choice of parameters (see ``fixed") induces poor results, in which the early stopping is quite needed to rescue performance (see ``fixed$^*$"). 
When the parameters are properly assigned, the early stopping would be helpful to reduce computation cost. 
Our learned policy leads to fast practical convergence as well as excellent performance, sometimes even outperforms the oracle policy tuned via inaccessible ground truth (in $\times2$ case). We note this is owing to the varying parameters across iterations  generated automatically in our algorithm, which yield  extra flexibility than constant parameters over iterations.
%Fig~(REF) shows the results 
Besides, we find the learned model-free policy produces suboptimal denoising strength/penalty parameters compared with our mixed model-free and model-based policy, and it also fails to learn early stopping behavior.

%To understand the usefulness of the early stopping mechanism, we also report the results of these polices with optimal early stopping. It should be noted some policies (\eg "fixed optimal"
%and "oracle") requires to access the ground truth to determine parameters, which is generally impractical in real testing scenarios. 
%Besides, to analyze whether the model-based RL benefits our algorithm. We compare our algorithm with the learned policy by model-free RL whose $\pi_2$ is optimized using the model-free DDPG algorithm \cite{lillicrap2015continuous}. Table \ref{tb:policy-comparison} exhibits the results of all policies: the bad choice of parameters (see ``fixed") induces poor results, in which the early stopping is quite needed to rescue performance (see ``fixed$^*$"). When the parameters are properly assigned, the early stopping would be helpful to reduce computation cost. 
%Moreover, our learned policy leads to fast practical convergence as well as excellent performance, sometimes even outperforms the oracle policy tuned via inaccessible ground truth (in $\times2$ case). We note this is owing to the varying parameters across iterations  generated automatically in our algorithm, which yield to extra flexibility than constant parameters over iterates.
%Besides, we find the learned model-free policy produces suboptimal denoising strength/penalty parameters and it fails to learn early stopping behavior as well.

\noindent {\bf Comparisons with state-of-the-arts.~} 
We compare our method against six state-of-the-art methods for CS-MRI, including the traditional optimization-based approaches (RecPF \cite{yang2010fast} and FCSA \cite{huang2010mri}), the PnP approaches (BM3D-MRI \cite{eksioglu2016decoupled} and IRCNN \cite{Zhang_2017_CVPR}), and the deep unrolling approaches (ADMMNet \cite{NIPS2016_6406} and ISTANet \cite{Zhang_2018_CVPR}).
To keep comparison fair, for each deep unrolling method, only single network is trained to tackle all the cases using the same dataset as ours. 
Table \ref{tb:CS-MRI} shows the method performance on  two set of medical images, \ie 7 widely used medical images (Medical7) \cite{huang2010mri} and 50 medical images from MICCAI 2013 grand challenge dataset\footnote{https://my.vanderbilt.edu/masi/}. 
The visual comparison can be found in Fig.~\ref{fig:comparison-cs-mri}. It can be seen that our approach significantly outperforms the state-of-the-art PnP method (IRCNN) by a large margin, especially under the difficult $\times8$ %acceleration
case. In the simple cases (\eg $\times2$), our algorithm only runs 5 iterations to arrive at the desirable performance, in contrast with 30 or 70 iterations required in IRCNN and BM3D-MRI respectively.

\subsection{Phase retrieval}
The goal of phase retrieval (PR) is to recover the underlying image from only the amplitude, or intensity of the output of a complex linear system.  Mathematically, PR can be defined as the problem of recovering a signal $x\in \mathbb{R}^{N}$ or $\mathbb{C}^{N}$ from measurement $y$ of the form $y^2 = \left| Ax \right|^2 + \omega$, 
% \begin{align}
%     y = \left| Ax \right| + \omega
% \end{align}
where the measurement matrix $A$ represents the forward operator of the system, and $\omega$ represents shot noise. We approximate it with $\omega \sim \mathcal{N} (0, \alpha^2|Ax|^2)$. The term $\alpha$ controls the sigma-to-noise ratio in this problem. 

We test algorithms with coded diffraction pattern (CDP) \cite{CandPhase2015pr}. Multiple measurements, with different random spatial modulator (SLM) patterns are recorded. We model the capture of four measurements using a phase-only SLM as \cite{metzler2018prdeep}. 
Each measurement operator can be mathematically described as $A_i = \mathcal{F}D_i, \quad i \in [1,2,3,4]$,
% \begin{align}
%     A = \left[ \substack{\mathcal{F} D_1 \\ \mathcal{F} D_2 \\ \mathcal{F} D_3 \\ \mathcal{F} D_4} \right]
% \end{align}
where $\mathcal{F}$ can be represented by the 2D Fourier transform 
and $D_i$ is diagonal matrices with nonzero elements drawn uniformly from the unit circle in the complex planes. 

We compare our method with three classic approaches (HIO \cite{Fienup1982Phase}, WF \cite{candes2014phase}, and DOLPHIn \cite{MairalDOLPHIn}) and three PnP approaches (SPAR \cite{katkovnik2017phase}, BM3D-prGAMP \cite{Metzler2016BM3D} and prDeep \cite{metzler2018prdeep}). 
Table \ref{tb:phase-retrieval} and Fig.~\ref{fig:comparison-phase-retrieval} summarize the results of all competing methods on twelve images used in \cite{metzler2018prdeep}. It can be seen that our method still leads to state-of-the-art performance in this nonlinear inverse problem, and produces cleaner and clearer results than other competing methods. 

\section{Conclusion}
In this work, we introduce RL into the PnP framework, yielding a novel tuning-free PnP proximal algorithm for a wide range of inverse imaging problems. 
We underline the main message of our approach \textit{the main strength of our proposed method is  the policy network, which can customize well-suited parameters for different images. }
%The key feature of our approach is the policy network, which can customize well-suited parameters for different images. 
Through numerical experiments, we demonstrate our learned policy often generates highly-effective parameters, which even often reaches to the comparable performance to the ``oracle" parameters tuned via the inaccessible ground truth.

\section*{Acknowledgements}
Authors gratefully acknowledge the financial support of  the CMIH and CCIMI University of Cambridge, and Graduate school of Beijing Institute of Technology.
 This work was also supported by the National Natural Science Foundation of China
 under Grant No. 61672096.
 
%\bibliography{egbib}
\bibliographystyle{icml2020}

\end{document}